\documentclass[aps,prl,notitlepage,twocolumn]{revtex4-1}

\usepackage{graphicx}
\usepackage{dcolumn}
\usepackage{bm}
 \usepackage{xcolor}

\newcommand{\x}{{\bf r}}
\newcommand{\K}{{\bf k}}

\begin{document}
\title{Quantum Bose-Bose droplets at a dimensional crossover}

\author{Pawe{\l} Zin$\,^{1,2}$, Maciej Pylak$\,^{2,3}$, Tomasz Wasak$\,^{2}$, Mariusz Gajda$\,^{3}$ and Zbigniew Idziaszek$\,^2$ }

\affiliation{\mbox{$^1$ National Centre for Nuclear Research, \mbox{ul.~Ho\.za~69}, PL-00-681 Warsaw, Poland} \\
\mbox{$^2$  Faculty of Physics, University of Warsaw, \mbox{ul.~Pasteura 5}, PL-02-093 Warsaw, Poland } \\
\mbox{$^3$ Institute of Physics, Polish Academy of Sciences, Aleja Lotnikow 32/46, PL-02-668 Warsaw, Poland}}

\date{\today}

\begin{abstract}
We study a liquid quantum droplets in a mixture of two-component Bose-Einstein condensates under a variable confinement introduced along one or two spatial dimensions.
Despite the atom-atom scattering has a three-dimensional character, discreetness of the available modes in the reduced dimension(s) strongly influences the
zero-point energy -- the Lee-Huang-Yang term. In a weakly interaction limit, it is the leading correction to the mean-field energy at the crossover from three
to two dimensions, or from three to one dimension. We analyze the properties of the droplets at the dimensional crossovers, and provide the demanding conditions for
accessing quasi-low dimensions. We predict new kinds of droplets which are formed only due to the quantum fluctuations
when the mean-field interaction vanishes. Our results pave the way for exploring new states of quantum matter, and are important for experiments with liquid quantum
droplets in reduced dimensions.
\end{abstract}

\maketitle


\textit{Introduction.}---Mixtures of two atomic Bose-Einstein condensates are the systems with a diverse spectrum of physical properties. The inter- and intra-species
interaction strengths, $g_{11}, g_{22}$, and $g_{12}$, respectively, are the key parameters defining their behavior. The energy density functional of a uniform mixture in
the mean field approximation is a quadratic form~\cite{stringari}:
\begin{equation}\label{MF}
\epsilon_\mathrm{MF} =\frac12 g_{11} n_1^2 + \frac12 g_{22} n_2^2 + g_{12} n_1 n_2, 
\end{equation}
where $n_1$ and $n_2$ are densities of the species.  The mixtures can be miscible if $|g_{12}| < \sqrt{g_{11}g_{22}}$, or immiscible if interspecies repulsion dominates,
$g_{12}> \sqrt{g_{11}g_{22}}$.  On the contrary, if inter-species attraction is strongly attractive, $g_{12} < - \sqrt{g_{11}g_{22}}$, a mixture collapses. Typically,
miscible mixtures have to be kept in external traps since, if left alone, they expand to minimize their energy.

The mean-field description overlooks existence of ultra-dilute quantum droplets ---  the exotic phases of the self-bound incompressible system of a two component Bose-Einstein
condensates (BECs), stabilized by quantum fluctuations~\cite{Petrov15}, and with densities orders of magnitude smaller than of ordinary liquids.

In a weakly interacting regime, the energy related to the quantum fluctuations is small, and, for a single-component BEC, is known as the Lee-Huang-Yang (LHY)
correction to the ground state energy of the system~\cite{Lee57}:
\begin{eqnarray}
\epsilon_\mathrm{LHY} =   \frac{128}{30\sqrt{\pi}}\, gn^2 \sqrt{na^3},
\end{eqnarray}
where $n$ is the density, the coupling strenght $g=4 \pi \hbar^2 a/m$, $a$ is the positive $s$-wave scattering length, and $m$ is the atomic mass.  The correction
$\epsilon_\mathrm{LHY}$ originates from a zero-point energy of the vacuum of Bogoliubov's quasiparticles. Since it depends on a higher power of the density, as compared
to the leading mean-field terms, its contribution to the energy is negligible in most circumstances.  However, for the Bose-Bose mixture, at the edge of the stability,
close to the collapse threshold, the mean-field energy vanishes, and the quantum fluctuations start to dominate.  As predicted in~\cite{Petrov15}, these fluctuations
contribute additional energy, called the LHY correction~\cite{Larsen63,Sacha08,Petrov15}, and stabilize the system and lead to the formation of quantum droplets.

Quantum droplets were first observed in Dysprosium and Erbium BECs~\cite{Kadau16, Ferrier16a,Ferrier16b,Schmitt16, Chomaz16}, in which the dipole-dipole interactions
between atoms is significant.  This anisotropic interaction, depending on the relative position of atoms and the orientation of their magnetic dipole moments, can be
attractive or repulsive.  The competition of attraction and repulsion, similarly to the two-component mixtures, might bring the system to the stability edge, making it
vulnerable to quantum fluctuations. The original scenario from Ref.~\cite{Petrov15}, was realized in the recent experiments with two-component Potassium
BECs~\cite{Cabrera17, Fattori17, Cheiney18}.

Quantum droplets can also exists in low-dimensional systems~\cite{Petrov16}. Due to the expected reduction of three-body losses, these droplets are of a great
experimental interest.  Such low-dimensional systems can be created by employing tight confinements in one or two spatial directions. However, tight externals potentials
significantly modify the excitation spectrum, and, in particular, the zero-point energy of the quasi-particles. Therefore, quantum droplets in reduced dimensions possess
different properties then those in three-dimensional (3D) space both for BEC mixtures~\cite{Petrov16} and for dipolar BECs~\cite{Mishra16}.

In experiments, the quasi-two-dimensional (quasi-2D) or quasi-one-dimensional (quasi-1D) regimes are obtained by a tight confinement introduced by external potential in
one or two directions.  The potential introduces an additional linear length scale $L$ of the tight confinement. This scale sets a lower limit on the excitation momentum
to $ \sim \hbar/L$ and minimal excitation energy $\varepsilon_0 = (\hbar^2/2m)(2\pi/L)^2$ in the confined direction(s).  If both thermal energy $k_BT$ as well as
characteristic interaction energy $\sim g_{11}n_1+g_{22}n_2$, are too small to allow for excitation in the tight direction(s), i.e. $\varepsilon_0 \gg k_BT$, and
$\varepsilon_0 \gg g_{11}n_1,g_{22}n_2$, i.e., $\varepsilon_0$ is the largest energy scale of the problem, then from a point of view of kinematics the system is
low-dimensional.

As shown in \cite{Petrov16}, the low-dimensional liquids are even more exotic then their 3D analogue. Three-dimensional droplets are formed when the mean-field approach
predicts a collapse of the system, i.e., interspecies attraction is sufficiently strong.  However, in lower dimensions, the two- and one-dimensional droplets can be
formed in an overall repulsive system, which liquefies while squeezed, and does not need any trapping potential in the not confined direction(s) anymore.

In our paper, we study the formation of droplets at the dimensional crossover from 3D to quasi-2D, and 3D to quasi-1D.  In this regime, which was not previously explored
in the literature, we find new kind of stable droplets which are formed only due to quantum fluctuations, when the mean-field interaction vanishes.  Our results are also
important for experiments for which the access to quasi-1D or quasi-2D regimes is demanding. Since such experiments are always performed in 3D under conditions of tight
confinement the kinematics may be low-dimensional to a large extent. The elimination of excitations, however, in the confined direction(s) is not complete, and the proper
description requires inclusion of corrections. In particular, we show that the access to quasi-1D is significantly more demanding than to quasi-2D, therefore, our results
are especially important for these experiments since in most circumstances only the crossover is accessible.

The conditions of the dimensional crossover may be reached by varying the trap geometries of the Bose-Bose mixture.  Both strongly prolate and oblate shapes of BECs can
be formed, and excitation energies in confined and extended direction(s) can be separated energetically, with a limited number of modes in confined direction(s) occupied
at low temperatures in the weakly interacting limit.  Such systems, with significantly varying spatial extensions in different directions, are in the region of
dimensional crossover.


\textit{Lee-Huang-Yang energy of a mixture in a box.}---The system we study is a two component mixture of interacting ultracold Bose gases in the ground state. The
mean-field energy density is given by Eq.~(\ref{MF}). Following the analysis presented in~\cite{Petrov15}, we consider the case when both intraspecies interactions are
repulsive, $g_{11}>0$, $g_{22}>0$, ($g_{11} \approx g_{22}$), while interspecies interaction is attractive, $g_{12}<0$. We also assume that the system is close to the
region of collapse, and, thus, the parameter $\delta g = g_{12} + \sqrt{g_{11}g_{22}}$ is small, i.e., $|\delta g| \ll g_{11},g_{22}$. The diagonal form of the mean-field
energy density reads:
\begin{eqnarray}
\epsilon_\mathrm{MF} = \lambda_-n_-^2 + \lambda_+n_+^2.\,
\end{eqnarray}
where the coefficients $\lambda_+ \simeq ({g_{11}+g_{22}})/2$, and $\lambda_- \simeq \delta g \sqrt{g_{11}g_{22}}/(g_{11}+g_{22})$.  In our regime $|\lambda_-| \ll
\lambda_+$, and thus the density $n_- = (n_1 \sqrt{g_{22}} + n_2 \sqrt{g_{11}})/(\sqrt{g_{11} + g_{22}})$ corresponds to a soft mode, while $n_+ = (n_1 \sqrt{g_{11}} -
n_2 \sqrt{g_{22}})/(\sqrt{g_{11} + g_{22}})$, is the density of a hard mode. Deviation of the latter from zero is energetically very costly, so we assume that in the
ground state the hard-mode density effectively vanishes, $n_+ = 0$. Consequently, the densities of both species are proportional to the density of the soft mode:
\begin{eqnarray}
n_- =  n_1 \sqrt{(g_{11}+g_{22})/g_{22}} = n_2 \sqrt{(g_{11}+g_{22})/g_{11}}. 
\end{eqnarray}
To further specify our system we assume that it is confined in a box, and periodic boundary conditions are imposed.  The standard LHY correction~\cite{Petrov15} in 3D is
evaluated under the assumption that all sides of the box have similar length. 

To find the LHY correction for the tightly confined system we have to consider the case when a one side of the box is much smaller than the others, $L_z \ll L_x \simeq
L_y$ (3D-2D crossover), or much larger (3D-1D crossover) $L_x \simeq L_y \ll L_z $, than remaining sides. These two configurations are considered separately below. We
denote the tight confinement extension by $L$ while a linear size of the box in perpendicular direction(s) by $L_{\perp}$.

At this stage, we do  not assume any particular geometry yet. The LHY energy density reads: 
\begin{equation} \label{LHY_mix}
 \frac{\varepsilon_0}{L^3}  e_\mathrm{LHY} = \lim_{r \rightarrow 0} \frac{\partial }{\partial r} \left( r \frac{1}{2V}\sum_\K e^{i\K\x} 
( \varepsilon_\K - A_\K) \right), 
\end{equation}
where $\varepsilon_\K = \sqrt{E_{k}^2 + 2 E_k(g_{11}n_1+g_{22}n_2) }$ and $ A_\K = E_k + g_{11}n_1+g_{22}n_2$, and $E_k= (\hbar^2k^2)/2m$. We extracted the prefactor
$\varepsilon_0/L^3$ to make $e_\mathrm{LHY}$ dimensionless. This form of the LHY energy results from a regularized pseudopotential~\cite{Lee57}, and it is equivalent to
the formula used in Ref.~\cite{Petrov15}, where the origin of the LHY term is attributed to the zero-point energy of the Bogoliubov vacuum.

In writing Eq.~(\ref{LHY_mix}), we made two approximations. First, we set $g_{12}^2 = g_{11}g_{22}$ which is consistent with the previous assumptions that the system is
about to collapse. This approximation is not a very restrictive one.  Second, we limit the analysis to mixtures of two species with equal masses only. Therefore, the
system we consider is, for instance, a mixture of atoms in two different internal spin states~\cite{Cabrera17,Cheiney18,Fattori17}.  The second approximations is quite
restrictive, however. We note that the LHY term is equal to the one of a single component Bose gas with effective $(gn)_\mathrm{eff} = g_{11}n_1+g_{22}n_2$.

The summation over discrete momentum states is essential to account for a tight confinement. If we substituted the summation over momenta with the integral, i.e., $1/V
\sum_{\bf k} \rightarrow \int {\rm d} {\bf k}/(2 \pi)^{3}$, we would recover the limit of an infinite box and the LHY energy of a Bose-Bose mixture in 3D
space~\cite{Petrov15}.

\textit{LHY energy at the 3D-2D crossover.}---Our main goal here is to find the LHY energy for a system confined in one spatial direction. In such a situation the
$z$-axis is a tight direction, i.e., $L=L_z$. Assuming that $L_x=L_y \rightarrow \infty$, in Eq.~(\ref{LHY_mix}) one has to substitute $\frac{1}{V}\sum_\K \rightarrow
\frac{1}{(2\pi)^2} \int {\rm d}^2 k_{\perp} \frac{1}{L} \sum_{k_z} $, and the LHY energy in quasi-2D takes the form:
\begin{equation} \label{LHY_2d}
e_\mathrm{LHY}^\mathrm{2d}(\xi)  = \lim_{r \rightarrow 0} \frac{\partial }{\partial r} \left(\! r   \frac{1}{2}\sum_{q_z} \int {\rm d}^2 q_{\perp} 
\, e^{i{\bf q} \x} \left( \varepsilon_{\bf q} - A_{\bf q} \right) \!\right),
\end{equation}
where $\xi=(g_{11} n_1 + g_{22} n_2)/\varepsilon_0$, ${\bf q} = ({\bf q}_\perp,q_z)$ and $q_z$, ${\bf q}_\perp$ are the integer dimensionless momenta: $q_z=(L/2\pi)k_z$,
and ${\bf q}_\perp=(L/2\pi) {\bf k}_{\perp}$.  Bogoliubov's energies expressed in the units of $\varepsilon_0$ are: $\varepsilon_{\bf q}= \sqrt{q^4 + 2 \xi q^2}$ and $
A_{\bf q}=q^2 + \xi$.  The ratio $\xi$ of the sum of mean field energies of both component to the excitation energy in the tight direction is the crucial parameter
characterizing the system.  We note that Eq.~(\ref{LHY_2d}) applies not only to a system at 3D-2D crossover, but also in the case of strongly oblate geometry, where the
characteristic spacing of kinetic momenta in the tighter direction is much larger than spacing in the perpendicular directions.  Then, the densely spaced momenta in the
perpendicular direction can be considered as continuous.

\begin{figure}[htb]
	\centering
		\includegraphics[width=0.49\textwidth]{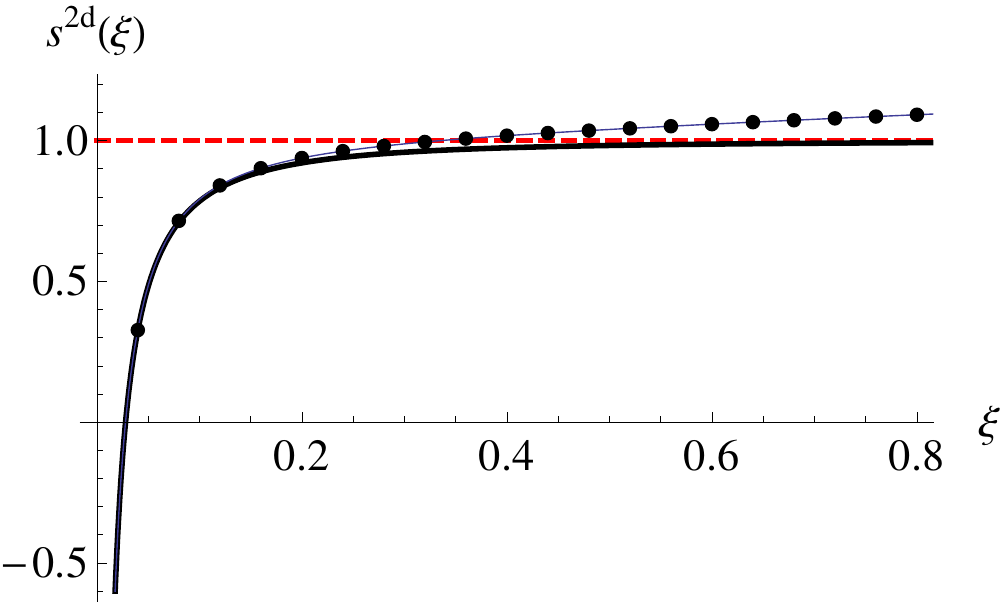}\\
		\caption{ The ratio $ s^\mathrm{2d}(\xi)= e_\mathrm{LHY}^\mathrm{2d}(\xi)/e_\mathrm{LHY}^\mathrm{3d}(\xi)$ as the function of $\xi$ given by the thick black
                  line.  The additional thin meshed curve is the same ratio but using the approximate formula for $e_\mathrm{LHY}^\mathrm{2d}(\xi)$ given by
                  Eq.~(\ref{e2d}). The red dashed horizontal line is the asymptotic 3D result.}
		\label{fig1}
\end{figure}

For small $\xi$, the result can be obtained analytically (see the Supplementary Material). The formula, derived for $\xi \ll 1$, is the following:
\begin{equation}\label{e2d}
e_\mathrm{LHY}^\mathrm{2d}(\xi) = \frac{\pi}{4} \xi^2 \left(   \log(\xi) + \log(2\pi^2) + \frac{1}{2}  +  \frac{\pi^2 \xi}{3} \right).
\end{equation}

We compare this expansion to the direct numerical evaluation of Eq.~(\ref{LHY_mix}).  In Fig.~\ref{fig1}, we plot the ratio $ s^\mathrm{2d}(\xi)=
e_\mathrm{LHY}^\mathrm{2d}(\xi)/e_\mathrm{LHY}^\mathrm{3d}(\xi)$, where the 3D LHY energy is $e_\mathrm{LHY}^\mathrm{3d}(\xi) = 16 \sqrt{2} \pi \xi^{5/2}/15$.  We also
plot there $s^\mathrm{2d}(\xi)$ but with $e_\mathrm{LHY}^\mathrm{2d}(\xi)$ taken from Eq.~(\ref{e2d}) (thin meshed curve).  The approximate expression for LHY term at
3D-2D crossover almost perfectly reproduces the numerical result for $\xi < 0.3$.  For larger values of $\xi$, the exact formula is in the perfect agreement with the 3D
expression.  The agreement between quasi-2D and 3D results for values of $\xi$ such small as $\xi=0.3$ is quite surprising because the 3D formula formally applies in the
limit $\xi \gg 1$.

\textit{Droplets at 3D-2D crossover.}---Neglecting the surface energy, which is well justified for large droplets where a bulk contribution dominates, the energy of the
homogeneous droplet of volume $V$ is equal to the sum of the mean-field term, $e_\mathrm{MF}$, and LHY correction, $e_\mathrm{LHY}^\mathrm{2d}$:
\begin{equation}
  \label{Ehom}
  E_\mathrm{hom} =  \frac{\varepsilon_0}{L^3} \left( e_\mathrm{MF}(\xi) + e_\mathrm{LHY}^\mathrm{2d}(\xi) \right) V,
\end{equation}
where $e_\mathrm{MF}(\xi) = \beta \xi^2$, and $\beta = \varepsilon_0 L^3 \delta g/\sqrt{g_{11}g_{22}} (\sqrt{g_{11}} +\sqrt{g_{22}})^2$ .

The droplet is stable in an empty space if its pressure vanishes, $ p= -({\rm d}/{\rm d}V) E_\mathrm{hom} =0 $.  Note, that $\xi$ is proportional to the density, i.e.,
${\rm d}\xi/{\rm d}V = - \xi/V$. The condition for the equilibrium density of droplets takes the form:
\begin{equation}
  \label{pzero}
  \left(\xi \frac{ \partial}{\partial \xi} -1 \right) \left( e_\mathrm{MF}(\xi) + e_\mathrm{LHY}^\mathrm{2d}(\xi)\right )  = 0.
\end{equation}

We now focus on the quasi-2D regime in which $e_\mathrm{LHY}^\mathrm{2d}(\xi)$ is given by Eq.~(\ref{e2d}) with the last term neglected.  Assuming for simplicity
$g_{11}=g_{22} = 4\pi \hbar^2 a/m$, which implies $n_1=n_2 =n$, the solution of Eq.~(\ref{pzero}) yields:
\begin{equation}
  \label{x0}
  \xi_0 = \frac{1}{2\pi^2}  e^{  -\frac{3}{2} -  \frac{ L \delta a}{2 a^2}}, 
\end{equation}
where we used $\delta g = 4 \pi \hbar^2 \delta a/m$. The above result leads to the following droplet density:
\begin{equation}
  \label{n2d}
  n=\frac{e^{-3/2}}{8 \pi} \frac{1}{a L^2}  e^{-\frac{L\delta a}{2a^2}}.
\end{equation}

To find the conditions for a quasi-2D system, we compare the droplet density obtained with Eq.~(\ref{e2d}) to the one given by Eq.~(\ref{n2d}).  We find that for $\xi
\lesssim 0.03$, the relative difference between the two results is smaller than $20\%$. We assume this condition defines the quasi-2D regime.  Therefore, to have a
quasi-2D system we need to have $\xi_0 \lesssim 0.03$ , and, from Eq.~(\ref{x0}), we find that
\begin{equation}
  \label{criterion}
  \frac{\delta a}{a} > - 4 \frac{a}{L} ,
\end{equation} 
According to Eq.~(\ref{criterion}), $\delta a$ can have arbitrary sign. Therefore, we arrive at the conclusion that droplets can be formed for the system with mean-field
energy corresponding to repulsive, weakly attractive, or even effectively vanishing interactions.  The last possibility was not discussed in the literature so far. It is
a droplet which is formed only due to quantum fluctuations.

Finally, let us compare our results for quasi-2D regime with the results from Ref.~\cite{Petrov16} for strictly 2D systems.  In the latter case, the LHY energy and
droplet densities are expressed in terms of 2D scattering length. Here, we consider the case $a \ll L$, i.e., the scattering has a 3D character. To compare the results,
we have to express the 2D scattering length, $a_\mathrm{2d}$, by the 3D one in the case of the box geometry analyzed in our paper. The scattering process in quasi-2D,
when the confinement in a tight direction is provided by a box of length $L$, is expressed by~\cite{Zin18}:
\begin{equation}
  \label{a2d3d}
  a_\mathrm{2d} = 2L e^{ - \gamma - \frac{L}{2a} }.
\end{equation} 
An analogous formula, in a situation when the tight confinement is provided by a harmonic potential, is given in \cite{shlyapnikov}. Inserting the above relation into
equations for 2D LHY energy density and droplet density of~\cite{Petrov16} we recover our results, given by Eqs.~(\ref{Ehom}) and~(\ref{x0}), together with
Eq.~(\ref{e2d}) with the last term neglected. This agreement provides an important and independent test of our approach.


\textit{LHY energy at the 3D-1D crossover.}---We now focus on the 3D-1D crossover regime where the LHY energy is:
\begin{equation}
  \label{LHY_1d}
  e_\mathrm{LHY}^\mathrm{1d}(\xi) \!=\!   \lim_{r \rightarrow 0} \frac{\partial }{\partial r} \left(\! r \frac{1}{2}\sum_{q_x,q_y} \int {\rm d} q_z 
  \, e^{i{\bf q} \x} \left( \varepsilon_{\bf q} \!-\! A_{\bf q} \right) \!\right),
\end{equation}
where $q_{x,y}$ are integers, $q_{x,y}=(L_{\perp}/2\pi)k_{x,y}$, and $q_z$ is a real-valued dimensionless momentum, $q_z=(L/2\pi) k_z$.  The Bogoliubov's energies
expressed in the units of $\varepsilon_0$ have the same form as in the 3D-2D system.  Similarly as before, we took $L_{\perp} \rightarrow \infty$, and, thus, we
substituted: $\frac{1}{V}\sum_\K \rightarrow \frac{1}{2\pi} \int d k \frac{1}{L^2} \sum_{k_x,k_y} $. For small $\xi$, we obtain (see the Supplementary Material)
\begin{equation}
  \label{e1d}
  e_\mathrm{LHY}^\mathrm{1d} =  - \frac{2\sqrt{2}}{3} \xi^{3/2} + c_2 \xi^2 + c_3 \xi^3,
\end{equation}
where $c_2 = \frac{1}{4} \left(  \int {\rm d} {\bf n} \,  1/n^2 - \sum_{n_y,n_z \neq 0} \int {\rm d} n_x   1/n^2    \right) 
\simeq 3.06$ and $c_3 = \frac{\pi}{8} \sum_{n_x,n_y \neq 0} (n_x^2+n_y^2)^{-3/2} \simeq 3.55 $.

\begin{figure}[htb]
  \centering
  \includegraphics[width=0.49\textwidth]{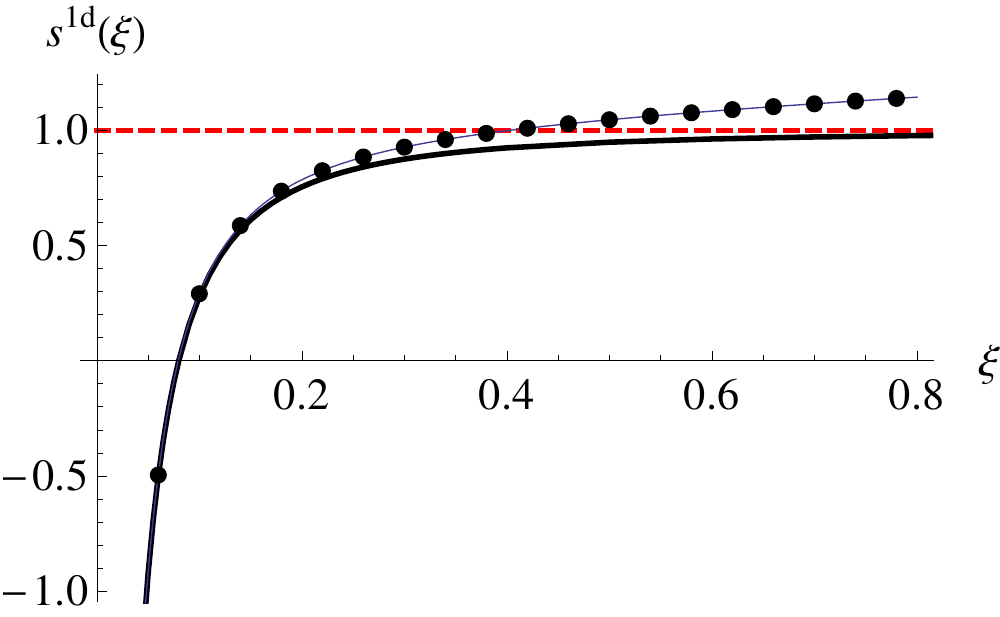}
  \caption{The ratio $ s^\mathrm{1d}(\xi)= e_\mathrm{LHY}^\mathrm{1d}(\xi)/e_\mathrm{LHY}^\mathrm{3d}(\xi)$ as the function of $\xi$ given by the thick black line.  
    The additional thin meshed curve is the same ratio but using the approximate formula for $e_\mathrm{LHY}^\mathrm{1d}(\xi)$ given by Eq.~(\ref{e1d}).}
  \label{fig2}
\end{figure}

In Fig.~\ref{fig2}, we plot the ratio $ s^\mathrm{1d}(\xi)= e_\mathrm{LHY}^\mathrm{1d}(\xi)/e_\mathrm{LHY}^\mathrm{3d}(\xi)$.  As before, the analytic approximate
expression for the 3D-1D LHY term almost perfectly matches the full numerical result for $\xi < 0.3$.  For larger values of $\xi$, the exact formula is close to the 3D
expression.

\textit{Droplets at 3D-1D crossover.}--- We now analyze the quasi-1D regime defined in the limit $\xi \ll 1$.  Including only the first term of Eq.~(\ref{e1d}), we find from
Eq.~(\ref{pzero}) that at the equilibrium $\xi_0 = (2/9)\beta^2 = (128/9\pi^2)a^4/(\delta a^2 L^2)$. The corresponding droplet density is
\begin{equation}\label{xi01}
  n = \frac{32 }{9\pi } \frac{a^3}{\delta a^2 L^4},
\end{equation}
where we assumed for simplicity $g_{11}=g_{22}= 4\pi \hbar^2 a/m$.  To find the condition for the validity of this formula, we compare it with the density of the droplet
using the full $e_\mathrm{LHY}^\mathrm{1d}(\xi)$ from Eq.~(\ref{e1d}). The relative density differs from the one given by Eq.~(\ref{xi01}) by $20\%$ for $\xi$
approximately equal to $0.0004$. Thus, for $\xi \lesssim 0.0004$, the formula for the density of the quasi-1D droplet is valid.  However such small value of $\xi$ is
probably out reach for current experiments.  As in the 2D-3D crossover we also find here droplets which exist for $\beta =0$.  Using Eqs.~(\ref{pzero}) and (\ref{e1d}) we
find their density corresponds to $\xi_0 \simeq 0.15$ which places such droplet far away from the quasi-1D regime and of course far away from the 3D system where such
droplet cannot exist.

We now compare our predictions to the 1D results obtained in~\cite{Petrov16}.  To this end, we have to express the 3D interaction parameter by the 1D coupling,
$g_\mathrm{1d}$.  From Ref.~\cite{olshanii}, we infer that, for $a/L \ll 1$, the $g_\mathrm{1d}$ can be obtained by averaging the 3D interaction over the density profile
in the tight directions, yielding $g_\mathrm{1d} = g/L^2 $. Using this relation, we obtain that in the quasi-1D regime the energy and equilibrium densities of the droplet
have the same form as given in~\cite{Petrov16,dopiska1}.


\textit{Validity of the approach.}---We now briefly discuss the validity of our results.  The Bogoliubov approach is valid as long as the LHY energy correction
$\frac{\varepsilon_0}{L^3} e_\mathrm{LHY} $ is much smaller than the characteristic mean-field energy density $gn^2$ (for simplicity we take $g_{11}=g_{22}=g$).  This
condition reads $\frac{\pi L}{2 a} \xi^2 \gg |e_\mathrm{LHY}|$. For both situations analyzed in our paper, $e_\mathrm{LHY}^\mathrm{1d,2d}$ is practically equal to
$e_\mathrm{LHY}^\mathrm{3d}$ for $\xi > 0.3$. Then, the condition is equivalent to the 3D condition, namely, $na^3 = \xi \frac{a^2}{L^2} \ll 1 $ , which we assume.  For
smaller values of $\xi$, we can use analytical formulas given in Eqs.~(\ref{e2d}) and~(\ref{e1d}), which lead to the condition $ |\log(\xi)| \ll \frac{2L}{a} $ in the
3D-2D and $\sqrt{\xi} \gg \frac{a}{L}$ in the 3D-1D case.

\textit{Conclusions.}---We analyzed the so far unexplored formation of quantum droplets in the Bose-Bose mixtures at dimensional crossover from 3D to 2D or 1D.  Under the
assumption, that the scattering processes are 3D, which happens when the spatial extent of a tight confinement $L$ is much larger than a 3D scattering length $a$, we have
found expressions for the beyond-mean field correction to the system energy.  These corrections generalize the Lee-Huang-Yang term as obtained for the 3D BEC.  We show
how this energy smoothly changes as a function of the parameter $\xi=(g_{11} n_1 + g_{22} n_2)/\varepsilon_0$.

The analysis of 3D-2D and 3D-1D crossovers revealed that the quasi-2D and quasi-1D regimes are accessed for values of $\xi\lesssim0.03$ and $\xi \lesssim 0.0004$,
respectively, which are much smaller than expected. The naive prediction suggesting that for $\xi < 1 $, the excitations in the confined directions are practically
frozen, and the system should be quasi-low-dimensional, does not work.  Counter-intuitively, we find that for $\xi > 0.3$ the LHY correction is practically equal to the
one obtained in the 3D case.

Our results provide the working parameters for the planned experiments, which aim at exploring low-dimensional formation of droplets in Bose-Bose mixtures.  The quasi-2D
regime, as compared to quasi-1D, is accessible for a broader range of~$\xi$, i.e., for $\xi \lesssim 0.03$, which is, however, still experimentally demanding.  The
quasi-1D regime is attained for a much smaller range of~$\xi$, i.e., $\xi\lesssim0.0004$, which poses a severe experimental constraint.  However, our work reveals that
yet unexplored 3D-1D crossover supports exotic droplets, different from both 3D and quasi-1D case, and formed only due to quantum fluctuations. Such droplets also exist
at the border of the quasi-2D regime. The results we present pave the way for exploring new states of matter in low-dimensional systems, in which quantum fluctuations
play the prominent role.

\begin{acknowledgments}
P.Z. and Z.I. acknowledge the support from the Polish National Science Center Grant No. 2015/17/B/ST2/00592.
M.P. and T.W. were supported by the Polish National Science Center Grant No. 2014/14/M/ST2/00015. 
M.G. acknowledges support from the (Polish) National Science Center Grant UMO-2017/25/B/ST2/01943 and from the EU Horizon 2020-FET QUIC 641122.
\end{acknowledgments}

\newpage
\setcounter{equation}{0}

\appendix

\section{SUPPLEMENTAL MATERIAL}

In this Supplemental Material we discuss the details of the derivation of Eqs. (7) and (15) from the main paper. We show how to properly handle the sums and integrals
in the Lee-Huang-Yang energy in order to arrive at a correct finite result.

\subsection*{Crossover from 3D to quasi-2D}

We start from the main equation for the Lee-Huang-Yang energy, i.e., Eq.~(6) from the main. We thus have:
\begin{eqnarray} \label{row1}
&& - \frac{2}{\xi^2} e_\mathrm{LHY}^\mathrm{2d}(\xi) 
\\ \nonumber
&& = \frac{\partial }{\partial r} \left (r 
\sum_{q_z} \int \mbox{d}^2 q_\perp  \frac{1}{\sqrt{q^2(q^2+2\xi)} + q^2+\xi}  e^{i{\bf q} \x } \right)\!\Bigg|_{r=0}. 
\end{eqnarray}
Now, we expand the right hand side in the power series in $\xi$, and we find
\begin{eqnarray*}
&& \sum_{q_z} \int \mbox{d}^2 q_\perp  \frac{1}{\sqrt{q^2(q^2+2\xi)} + q^2+\xi}  e^{i{\bf q} \x } 
\\
&& \simeq \int \mbox{d}^2 q_\perp  \frac{1}{\sqrt{q_\perp^2(q_\perp^2+2\xi)} + q_\perp^2+\xi}  e^{i{\bf q} {\bf r}}
\\
&&
+ \sum_{q_z \neq 0} \int \mbox{d}^2 q_\perp \, \frac{1}{2q^2} e^{i{\bf q} \x}
- \xi \sum_{q_z \neq 0} \int \mbox{d}^2 q_\perp \, \frac{1}{2q^4} e^{i{\bf q} \x}.
\end{eqnarray*}

As we show in details below, in the limit $r\to 0$, we find that the first two terms are of the following form:
\begin{eqnarray} \nonumber
&& \int \mbox{d}^2 q_\perp  \frac{e^{i{\bf q} {\bf r}} }{\sqrt{q_\perp^2(q_\perp^2+2\xi)} + q_\perp^2+\xi}  
+ \sum_{q_z \neq 0} \int \mbox{d}^2 q_\perp \, \frac{ e^{i{\bf q} \x}}{2q^2}  
\\ \nonumber
\\ \label{row2}
&& =
\frac{\pi^2}{r}  - \frac{\pi}{2} \left( \log \xi + \frac{1}{2} + \log(2\pi^2) \right),
\end{eqnarray}
whereas the third term is given by
\begin{eqnarray*}
\sum_{q_z \neq 0} \int \mbox{d}^2 q_\perp \, \frac{ e^{i{\bf q} \x}}{2q^4} 
\simeq \sum_{q_z \neq 0} \int \mbox{d}^2 q_\perp \, \frac{1}{2q^4}
 = \sum_{n_z > 0} \frac{\pi}{n_z^2}= \frac{\pi^3}{6}.
\end{eqnarray*}
Inserting these expressions into Eq.~(\ref{row1}), we find
\begin{eqnarray*}
e_\mathrm{LHY}^\mathrm{2d}(\xi) = \frac{\pi}{4} \xi^2 \left( \log \xi + \frac{1}{2} + \log(2\pi^2)  + \frac{\pi^2}{3} \xi \right),
\end{eqnarray*}
which is Eq.~(7) of the main paper.

Now, we show in details how to derive Eq.~(\ref{row2}).
In order to simplify the calculation, we take $z=0$ in the vector $\x=(x,y,z)$, which gives $r = \sqrt{x^2+y^2}$.

We rewrite the first term from the left-hand side of Eq.(\ref{row2}) in cylindrical coordinates:
\begin{eqnarray*}
&& \int \mbox{d}^2 q_\perp  \frac{e^{i{\bf q} {\bf r}} }{\sqrt{q_\perp^2(q_\perp^2+2\xi)} + q_\perp^2+\xi}  
\\
&&  = 2\pi \int_0^\infty q_\perp \mbox{d} q_\perp \,  \frac{J_0(q_\perp r)}{\sqrt{q_\perp^2(q_\perp^2+2\xi)} +q_\perp^2+\xi}.
\end{eqnarray*}
The change of variable into $n = q_\perp r$ yields:
\begin{eqnarray*}
&&
= 2\pi \int_0^\infty n \mbox{d} n  J_0(n) \frac{1}{\sqrt{n^2(n^2+2\xi r^2)} + n^2+\xi r^2}
\\
&& \simeq 2\pi \int_0^{n_0} n \mbox{d} n   \frac{1}{\sqrt{n^2(n^2+2\xi r^2)} + n^2+\xi r^2}
\\
&&\quad + 2\pi \int_{n_0}^\infty  \mbox{d} n  \, \frac{J_0(n)}{2n}. 
\end{eqnarray*}
Here, we approximated $J_0(n)\approx 1$ in the first integral on right-hand side, and neglected $\xi r^2$ terms in the denominator in the second integral since the main
contribution comes from large $n$. These approximations are valid as long as $n_0 \ll 1$. In these limit, the integrals can be evaluated:
\begin{equation}
  \int_0^{n_0} n \mbox{d} n   \frac{ 2\pi}{\sqrt{n^2(n^2+2\xi r^2)} + n^2+\xi r^2} \simeq \frac{\pi}{2} \log \left(\frac{2n_0^2}{\xi r^2} \right)  - \frac{\pi}{4}
\end{equation}
and
\begin{equation}
  2\pi \int_{n_0}^\infty  \mbox{d} n  \frac{J_0(n)}{2n} \simeq  -\pi \left( \log(n_0) - \log(2) + \gamma \right),
\end{equation}
where $\gamma$ denotes the Euler's constant.
As a result, we obtain
\begin{eqnarray} \nonumber
&& \int \mbox{d}^2 q_\perp  \frac{e^{i{\bf q} {\bf r}} }{\sqrt{q_\perp^2(q_\perp^2+2\xi)} + q_\perp^2+\xi} 
\\ \label{w1}
&& = \frac{3\pi}{2} \log(2) - \pi\gamma  - \frac{\pi}{2} \log(\xi r^2)  - \frac{\pi}{4}.
\end{eqnarray}

Let us now analyze the second integral in Eq.~(\ref{row2}):
\begin{eqnarray*}
&& \sum_{q_z \neq 0} \int \mbox{d}^2 q_\perp \, \frac{ e^{i{\bf q} \x}}{2q^2} 
 = \pi \sum_{q_z \neq 0} \int_0^\infty q_\perp \mbox{d}q_\perp \, \frac{J_0(q_\perp r)}{q_\perp^2+q_z^2}
\\
&&
 = \pi \int_0^\infty q_\perp \mbox{d}q_\perp \, J_0(q_\perp r) \frac{\pi q_\perp \coth(q_\perp \pi) -1 }{q_\perp^2}.
\end{eqnarray*}
Now, we change the variables into $n = q_\perp r$, and take $\epsilon$ as the lower limit of the integral. 
The right-hand side is:
\begin{eqnarray*}
 \frac{\pi}{r}\int_{\epsilon }^\infty \mbox{d} n \,   J_0(n)\coth \left( \frac{n \pi}{r} \right)  
- \int_{\epsilon }^\infty \mbox{d} n \, \frac{J_0(n)}{n}.
\end{eqnarray*}
Now, the second term in this expression is:
\begin{eqnarray*}
 - \int_{\epsilon }^\infty \mbox{d} n \, \frac{J_0(n)}{n} \simeq \log(\epsilon/2) + \gamma,
\end{eqnarray*}
whereas the first gives:
\begin{eqnarray*}
&&\frac{\pi}{r}\int_{\epsilon }^\infty \mbox{d} n \,  J_0(n) \coth \left( \frac{n \pi}{r} \right)  
\simeq \frac{\pi}{r}\int_{\epsilon }^{n_0} \mbox{d} n \,  \coth \left( \frac{n \pi}{r} \right)
\\
&&    
 +  \frac{\pi}{r} \int_{n_0}^\infty \mbox{d} n \,   J_0(n)
 \simeq \log \left( \frac{r}{2\pi \epsilon} \right)   + \frac{\pi n_0}{r} + \frac{\pi}{r}(1-n_0),
 \end{eqnarray*}
where $ \epsilon \ll n_0 \ll 1$ and $n_0 \gg r$.
Finally, from these results altogether, we obtain
\begin{equation} \label{w2}
\sum_{q_z \neq 0} \int \mbox{d}^2 q_\perp \, \frac{ e^{i{\bf q} \x}}{2q^2} = \frac{\pi^2}{r} + \pi \log \left( \frac{r}{4\pi} \right) + \pi \gamma.
\end{equation}
Therefore, the sum of the expressions from Eqs.~(\ref{w1}) and~(\ref{w2}) recovers Eq.~(\ref{row2}).

\subsection*{Crossover from 3D to quasi-1D}

From Eq.~(14) of the main paper, we find
\begin{eqnarray} \label{row21}
&& - \frac{2}{\xi^2} e_\mathrm{LHY}^\mathrm{1d}(\xi) 
\\ \nonumber
&& = \frac{\partial }{\partial r} \left (r 
\sum_{q_x,q_y} \int \mbox{d} q_z  \frac{1}{\sqrt{q^2(q^2+2\xi)} + q^2+\xi}  e^{i{\bf q} \x } \right). 
\end{eqnarray}
As before, we this expression in powers of $\xi$, and we get:
\begin{eqnarray*}
&& \sum_{q_x,q_y} \int \mbox{d} q_z  \frac{e^{i{\bf q} \x } }{\sqrt{q^2(q^2+2\xi)} + q^2+\xi}   
\\
&& \simeq \int \mbox{d} q_z  \frac{ e^{i{\bf q} {\bf r}} }{\sqrt{q_z^2(q_z^2+2\xi)} + q_z^2+\xi}  
\\
&&
+ \sum_{q_x,q_y}{}\!' \int \mbox{d} q_z \, \frac{ e^{i{\bf q} \x}}{2q^2} 
- \xi \sum_{q_x,q_y}{}\!' \int \mbox{d} q_z \, \frac{ e^{i{\bf q} \x}}{2q^4},
\end{eqnarray*}
where $\sum_{q_x,q_y}'$ denotes the sum without the $q_x=q_y=0$ term.
In the limit $r\to 0$, the first term is:
\begin{eqnarray*}
&& \int \mbox{d} q_z  \frac{ e^{i{\bf q} {\bf r}}}{\sqrt{q_z^2(q_z^2+2\xi)} + q_z^2+\xi}  
\\
&& \simeq \int \mbox{d} q_z  \frac{1}{\sqrt{q_z^2(q_z^2+2\xi)} + q_z^2+\xi}
=  \frac{4\sqrt{2}}{3 \sqrt{\xi}},
\end{eqnarray*}
the second term is:
\begin{eqnarray*}
&& \sum_{q_x,q_y}{}\!' \int \mbox{d} q_z \, \frac{ e^{i{\bf q} \x} }{2q^4} 
\simeq \sum_{q_x,q_y}{}\!' \int \mbox{d} q_z \, \frac{ 1 }{2q^4}
\\
&& =  \frac{\pi}{4} \sum_{q_x,q_y}{}\!' \frac{1}{(q_x^2+q_y^2)^{3/2}} = 2c_3,
\end{eqnarray*}
and the third is:
\begin{equation}\label{c2}
 \sum_{q_x,q_y}{}\!' \int \mbox{d} q_z \, \frac{ e^{i{\bf q} \x}}{2q^2}
\simeq \frac{\pi^2}{r} - 2c_2,
\end{equation}
where $c_2$ is a constant given by
\begin{eqnarray} 
  \nonumber
  -\frac{4}{\pi}c_2 &=& \lim_{q_c \rightarrow \infty} \left( \sum_{|q_x|,|q_y| \leq q_c}{}\!\!\!\!\!\!\!'\quad \frac{1}{\sqrt{q_x^2+q_y^2}} \right.
  \\ 
  & & 
  \left.
  - \int_{-q_c-1/2}^{q_c+1/2} 
  \mbox{d}q_x   \int_{-q_c-1/2}^{q_c+1/2}  \mbox{d}q_y \, \frac{1}{\sqrt{q_x^2+q_y^2}} \right)\label{c22}.
\end{eqnarray} 
In these expression, the sum $\sum'$ means that we exclude the term with $q_x=q_y=0$. 
Collecting the terms altogether, we recover Eq.~(15) of the main paper.

Below, we show in details how to derive Eqs.~(\ref{c2}) and~(\ref{c22}).
To start, we assume $\x = (r,0,0)$ for simplicity, and, then, we obtain
\begin{eqnarray*}
\sum_{q_x,q_y}{}\!' \int \mbox{d} q_z \, \frac{ e^{i q_x r}}{2q^2}
= \sum_{q_x,q_y}{}\!' \frac{\pi}{2\sqrt{q_x^2+q_y^2}} e^{ -r\sqrt{q_x^2+q_y^2} }.
\end{eqnarray*}
In the limit $r \to 0$, the exponent is important for ensuring the convergence for large $\sqrt{q_x^2+q_y^2}$.  Therefore, to proceed, we divide the region of summation
into two parts $A$ and $B$: $A$ is the square region in which $|q_x|,|q_y| \leq q_c$, and $B$ is the rest. In the region $A$, we neglect the exponent, whereas in region $B$
we approximate the sum by the integral.  As a result, we obtain
\begin{eqnarray*}
&&\sum_{q_x,q_y}{'} \frac{1}{\sqrt{q_x^2+q_y^2}} e^{ -r\sqrt{q_x^2+q_y^2}} \\
&& \simeq \sum_{|q_x|,|q_y| \leq q_c}\!\!\!\!\!\!{'}\quad \frac{1}{\sqrt{q_x^2+q_y^2}} 
+ \int_B \mbox{d} q_x \mbox{d} q_y \frac{e^{ -r\sqrt{q_x^2+q_y^2}}}{\sqrt{q_x^2+q_y^2}},
\end{eqnarray*}
where $\int_B$ denotes the integral over the region $B$.
Now, we rewrite this integral as $\int_B = \int - \int_A$.
Here, it is important to observe that
\begin{eqnarray*}
 \int_A \mbox{d} q_x \mbox{d} q_y = \int_{-q_c-1/2}^{q_c+1/2} \mbox{d} q_x \int_{-q_c-1/2}^{q_c+1/2} \mbox{d} q_y.
\end{eqnarray*}
The shift in the boundaries by $1/2$ comes from the fact that summed element with $q_x,q_y$ is replaced in the integral by a square
of unit length with $q_x,q_y$ located at the center of the square.
As a result, we end up with
\begin{eqnarray*}
 \sum_{q_x,q_y}{'} \int \mbox{d} q_z \, \frac{ e^{i q_x r}}{2q^2} \simeq \int \mbox{d} {\bf q} \, \frac{ e^{i q_x r}}{2q^2}
 -2c_2 = \frac{\pi^2}{r} - 2c_2.
\end{eqnarray*}
In the above we notice that the limit of integration is $q_c+1/2$. We stress here the presence of the shift by $1/2$ in the integral's boundaries.
If the shift is neglected (as it would be done in approximated treatment) $c_2$ changes significantly,
influencing the properties of the system.


\begin{thebibliography}{99}
\bibitem{dopiska1} The coupling constant  $g_{1d}$ is related to $g$ by $g_{1d} \simeq g(1+C a/L)$, where $C$ is a constant of the order of unity.
It turns out that the correction to the energy density,  given by the $g C a/L$ term in the coupling constant, is negligible in the limit $\xi \ll 1$.
\end{thebibliography}
\end{document}